\documentclass[12pt,a4paper]{article}

\usepackage{amsmath}
\usepackage{amsthm,amssymb}
\usepackage{graphicx} 

\newcommand{\Ref}[1]{(\ref{#1})}

\newcommand{\beq}{\begin{equation}}
\newcommand{\eeq}{\end{equation}}

\newcommand{\pdiff}[2]{\frac{\partial#1}{\partial#2}}
\newcommand{\ppdiff}[3]{\frac{\partial^{#1} #2}{\partial {#3}^#1}}

\numberwithin{equation}{section}

\begin{document}

\title{Exact solution of two friendly  walks above a  sticky wall with single and double interactions}
\author{Aleksander L Owczarek$^1$ and Andrew Rechnitzer$^2$ and Thomas Wong$^2$\\[1ex]
  \footnotesize
  \begin{minipage}{9cm}
  $^1$Department of Mathematics and Statistics,\\
    The University of Melbourne, Victoria~3010, Australia.\\
    \texttt{owczarek@unimelb.edu.au}\\[1ex]
    $^2$Department of Mathematics,\\
  University of British Columbia,\\
  Vancouver V6T 1Z2, British Columbia, Canada.\\
  \texttt{andrewr@math.ubc.ca,twong@math.ubc.ca}
    \end{minipage}
}

\maketitle  

\begin{abstract}
We find, and analyse, the exact solution of two friendly directed walks, 
modelling polymers, which interact with a wall via contact interactions. We
specifically consider two walks that begin and end together so as to imitate a
polygon. We examine a  general model in which a separate interaction parameter is
assigned to configurations where both polymers touch
the wall simultaneously, and investigate the effect this parameter has on the
integrability of the problem. We find an exact solution of the generating
function of the model, and provide a full analysis of the phase diagram that
admits three phases with one first-order and two second-order transition lines
between these phases. We argue that one physically realisable  model would see two
phase transitions as the temperature is lowered. 
\end{abstract}

\section{Introduction}

The adsorption of polymers on a sticky wall, and confined between two walls, has
been the subject of continued interest
\cite{privman1988c-a,debell1993a-a,janse2000a-a,mishra2003a-a,brak2005a-:a,
janse2005a-:a,martin2007a-:a,owczarek2008c-:a,alvarez2008a-a,owczarek2009b-:a,
janse2010a-a,rychlewski2011a-a}. This has been in part due to the advent of
experimental techniques able to micro-manipulate single polymers
\cite{svoboda1994a-a,ashkin1997a-a,strick2001a-a} and the connection to
modelling DNA denaturation
\cite{essevaz-roulet1997a-a,lubensky2000a-a,lubensky2002a-a,orlandini2001a-a,
marenduzzo2001a-a,marenduzzo2002a-a,marenduzzo2003a-a,marenduzzo2009a-a}.

Consider a polymer in a dilute solution of \emph{good} solvent, so that it is in
a swollen state \cite{gennes1979a-a}. If such a polymer is then attached to a wall at one end
the rest of the polymer drifts away due to entropic repulsion. On the other hand, if  the wall
has an attractive contact potential, so that it becomes `sticky' to the
monomers, the polymer can be made to stay close to the wall by a sufficiently
strong potential, or for low enough temperatures. The  phase transition
between these two states is the \emph{adsorption}
transition. The high temperature state is \emph{desorbed} while the low
temperature state is \emph{adsorbed}. This pure adsorption transition has been
well studied \cite{privman1988c-a, debell1993a-a, hegger1994a-a, janse2000a-a,
janse2004a-a} exactly and numerically, and has been demonstrated to be
second-order.

There has been recent interest \cite{alvarez2008a-a} in ring polymers, modelled by 
self-avoiding polygons,  being adsorbed onto the walls of a two-dimensional
slit. In that work, models in which both sides of the polygon could interact
with each of the walls were considered. This provides us with one of the motivations for the model
here, where we consider two directed walks that begin and end together, so
forming a polygon. We consider directed walks because they often admit exact
solutions, while the more realistic self-avoiding walks do not. Moreover, we
consider such a pair of walks interacting with a sticky wall, allowing different
interactions when one or both walks are near the wall. To allow for a simple
realisation of the model we consider so-called friendly directed walks (rather
than the ubiquitous vicious walks) where the two walks may share edges of the
lattice. However, we do not allow the walks to cross and so there is always an
\emph{upper} walk/polymer and a \emph{lower} walk/polymer.

Other physical motivations for two-walk models have appeared in the literature.
In particular, one may model DNA-denaturation in this way --- for example the
Poland-Scheraga models \cite{poland1970theory, richard2004poland}. It would be interesting to see if the
techniques described below could be used to find exact solutions of the DNA
unzipping transition in the presence of an adsorbing wall, such as that
discussed in \cite{kapri2009can}. To do so we would add a contact interaction
between the two walks, rather than the double-visit interaction discussed here.
A manuscript on this topic is in preparation \cite{AAR2012}.

As we investigated the model another motivation for its interest became
apparent; the full two parameter model is not amenable to one of the standard
methods of solving multiple walk models. The Lindstr\"om-Gessel-Viennot lemma
\cite{gessel1985a-a, lindstrom1973a-a} (which was also considered earlier in a
probabilistic context by Karlin and McGregor \cite{karlin1959a-a}) decomposes
the solution of models of multiple vicious walks into combinations of single
walk problems. The lemma implies that the generating function of a multiple walk
model would be governed by a D-finite (Differentiably-Finite) function,
but the solution of our model is not D-finite. Despite this, we
are still able to solve the model.

We have solved our model in two ways. Firstly, we use the obstinate
kernel method (see \cite{bousquet2010a-a} for an overview of the technique) to
give a formal solution of a functional equation as a constant term formula. This
constant term can then be evaluated explicitly. Secondly, we also use a
`primitive piece' decomposition that allows us to give an explicit solution
in terms of hypergeometric-type sums.

Our solution allows us to fully analyse the model and we find a rich phase
diagram. In particular, we find three phases that meet at a special point. There are three phase
boundaries; two are second-order and one is first-order. Intriguingly, we find
that arguably the most physically realisable one-parameter case of our model would have
two phase transitions on lowering the temperature.

In the next section (Section~\ref{sec:model}), we formally define our model. In
Section~\ref{sec:funeqn}, we formulate a functional equation obeyed by an
extended  generating function and provide a `constant term' solution in
Section~\ref{sec:solfuneqn} via the obstinate kernel method. We provide an
alternate explicit solution in Section~\ref{sec:altsol} that illustrates an
underlying structure in the solution which arises combinatorially. In
Section~\ref{sec:analpd}, we analyse the phase structure and phase transitions
of the model. In the final section (Section~\ref{sec:disc}) we discuss the functional nature of the solution and summarise our
results by recasting them in terms of some physical parameters of a family of
single parameter models.

\section{Model}
\label{sec:model}
We consider a pair of directed walks above a wall on the upper half-plane of the
square lattice, taking steps $(1,1)$ or $(1,-1)$. These walks may touch but not
cross; such walks are sometimes called \emph{friendly} walks. Further, we
consider those pairs of walks that begin at the point $(0,0)$ and have equal
length. Let $\varphi$ be a pair of such walks and the set of all such walks be $\Omega$. 
We define $|\varphi|$ to be the length of the walks. 

To these configurations we add an energy $-\varepsilon_a$ for visits of the
bottom walk only (\emph{single visits}) to the wall, and an energy
$-\varepsilon_d$ when both walks visit a site on the wall simultaneously
(\emph{double visits}), excluding the first vertex of the walks. The number of
\emph{single visits}  to the wall  will be denoted $m_a({\varphi})$, while the
number of \emph{double visits} will be denoted $m_d({\varphi})$. 

\begin{figure}[h]
\begin{center}
 \includegraphics[width=14cm]{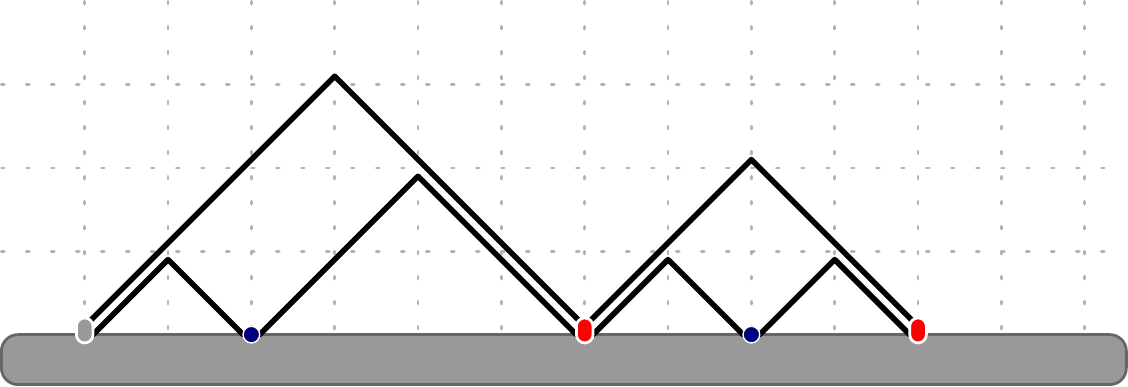}
\end{center}
 \caption{Two directed walks of length 10 of our model that begin and end on
the surface. There are two single and two double visits marked. The left-most
(start) vertex of the two walks on the wall is not counted as a double visit.}
\label{model}
\end{figure}

Later in the paper we will specialise to those configurations,
$\widehat{\varphi}$ in which both walks start and end on the wall. Since every
such configuration has at least one double visit (the final vertex), we
have $m_d(\widehat{\varphi})\geq 1$. The partition function for our model is
\begin{equation}
Z_n(a,d) = \sum_{\widehat{\varphi}\, \ni\, |\widehat{\varphi}|=n} 
  e^{\left(m_a(\widehat{\varphi}) \cdot \varepsilon_a +
m_d(\widehat{\varphi}) \cdot \varepsilon_d \right)/k_B T}
\end{equation}
where $T$ is the temperature and $k_B$ the Boltzmann constant, and associated
Boltzmann weights are denoted $a=e^{\varepsilon_a/k_B T}$ and
$d=e^{\varepsilon_d/k_B T}$. The thermodynamic reduced free energy of our model
is given in the usual fashion as
\begin{equation}
\kappa(a,d) = \lim_{n \rightarrow \infty} n^{-1} \log\left(Z_n(a,d)\right).
\end{equation}
A configuration of length $10$ in our model with single and double visits marked
appears in Figure~\ref{model}.

To find the free energy we will instead solve for the generating function
\begin{equation}
G(a,d;z) = \sum_{n=0}^\infty Z_n(a,d) z^n.
\end{equation}
The radius of convergence of the generating function $z_c(a,d)$ is directly
related to the free energy via 
\begin{equation}
\kappa(a,d) = \log(z_c(a,d)^{-1}).
\end{equation}

\section{Functional Equations}
\label{sec:funeqn}
To find $G$, we consider walks  $\varphi$ in the larger set, where each
walk can end at any possible height.  Let us first consider $a=d=1$. In this
case we construct the expanded generating function
\begin{align}
F(r,s;z) \equiv F(r,s) 
&= \sum_{\varphi \in \Omega} z^{|\varphi|}
r^{\lfloor\varphi\rfloor} s^{\lceil \varphi \rceil/2},
\end{align}
where $z$ is conjugate to the length ${|\varphi|}$ of the walk, $r$ is conjugate to the
distance ${\lfloor\varphi\rfloor}$ of the bottom walk from the wall and $s$ is conjugate to \emph{half}
the distance ${\lceil \varphi \rceil}$ between the final vertices of the two walks. Since the distance
between the endpoints of the walks changes by $0$ or $\pm 2$ with each step, and
the endpoints start together, it is always an even number. Further, we let
$[r^js^k]F(r,s)$ denote the coefficient of $r^js^k$ in the generating function
$F(r,s)$. We use $[r^j]F(r,s)$ to denote the coefficient of $r^j$ in
$F(r,s)$ which is a function of $s$ and similarly $[s^k]F(r,s)$ gives a
function of $r$.

%In order to obtain the full functional equation, we now add the two interaction
%parameters $a$ and $d$, with $a$ conjugate to the number of contacts between the
%bottom walk and the wall and $d$ be conjugate to the number of contacts between
%both walks and the bottom wall as described above. \textbf{Do we need this
%paragraph? --- I think not.}

Let us now return to general $a$ and $d$. All pairs of walks can then be built using a standard column-by-column
construction. Translating this into its action on the generating function gives
the following functional equation
\begin{align}
F(r,s) =& 1 + z \left(r + \frac{1}{r} + \frac{s}{r}+\frac{r}{s}\right) \cdot
F(r,s) \nonumber \\
& - z\left(\frac{1}{r}+\frac{s}{r}\right)\cdot [r^0]F(r,s) - z\frac{r}{s}\cdot
[s^0]F(r,s) \nonumber\\
& + z(a-1)(1+s)\cdot [r^1]F(r,s) + z(d-a) \cdot [r^1s^0]F(r,s).
\end{align}

\begin{figure}[h]
\begin{center}
 \includegraphics[width=14cm]{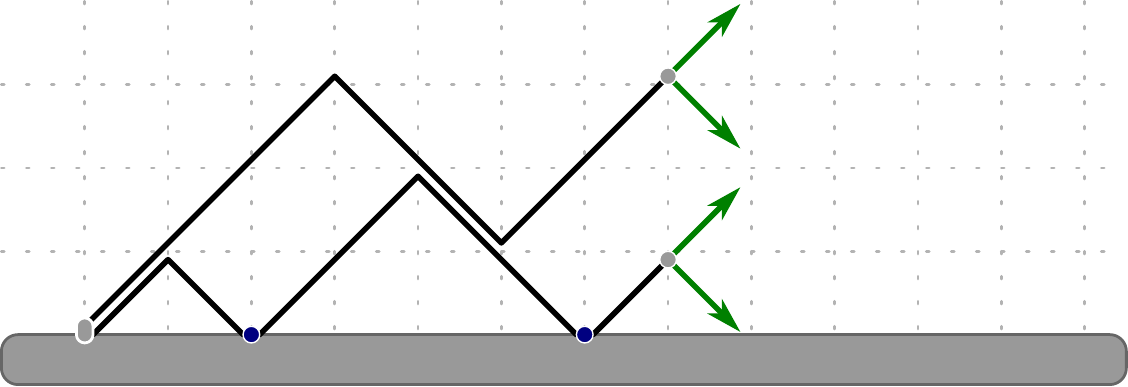}
\end{center}
\caption{Adding steps to the walks when the walks are away from the wall. There are four possibilities.}
\label{steps}
\end{figure}
We explain each of the terms in this equation. 
\begin{itemize}
\item The trivial pair of walks of length $0$ gives the initial $1$ in the
functional equation.
\item Every pair of walks may be extended by appending directed steps to their
endpoints in four different ways (see Figure~\ref{steps}).
\begin{center}
\begin{tabular}{||c|c|c||}
\hline
Top walk & Bottom walk & Generating Function\\
\hline
$(1,1)$ & $(1,1)$ & $r\cdot F(r,s)$\\
$(1,1)$ & $(1,-1)$ & $\frac{s}{r}\cdot F(r,s)$\\
$(1,-1)$ & $(1,1)$ & $\frac{r}{s}\cdot F(r,s)$\\
$(1,-1)$ & $(1,-1)$ & $\frac{1}{r}\cdot F(r,s)$\\
\hline
\end{tabular}
\end{center}
\begin{figure}[h]
\begin{center}
 \includegraphics[width=14cm]{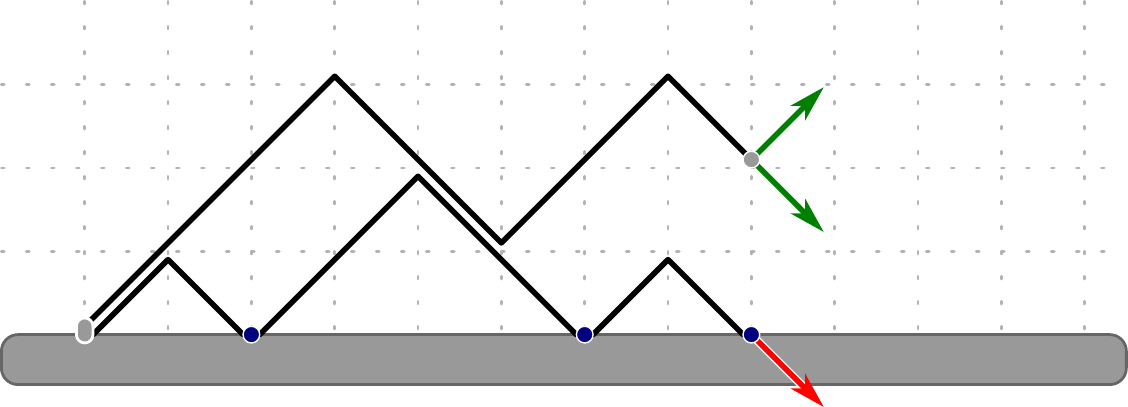}
\end{center}
\caption{The first boundary term in the functional equation removes the
contribution from the walks that are produced by appending a SE step to the
bottom walk when its endpoint is on the wall.}
\label{bndry_term1}
\end{figure}
\item Appending steps in this way may result in the bottom walk stepping below
the wall (Figure~\ref{bndry_term1}). Thus, when the bottom walk is at the wall,
we cannot append any steps that will decrease the height of the bottom walk.
These forbidden configurations are counted by
\begin{align}
 z\left(\frac{1}{r}+\frac{s}{r}\right)\cdot [r^0]F(r,s) 
& = z\left(\frac{1}{r}+\frac{s}{r}\right)\cdot F(0,s).
\end{align}
\begin{figure}[h]
\begin{center}
 \includegraphics[width=14cm]{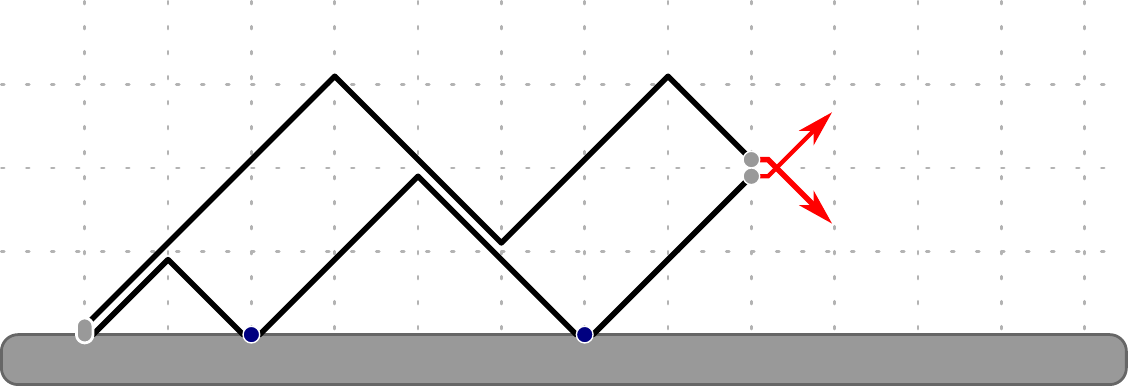}
\end{center}
\caption{The second boundary term in the functional equation removes the
contribution of walks that cross. Such configurations are produced when one
appends steps to walks that end at the same vertex as shown.}
\label{bndry_term2}
\end{figure}
\item Similarly, at no time can the top walk pass below the bottom walk
(Figure~\ref{bndry_term2}). Thus, if the two walks are touching, we forbid the
distance between them to decrease. These configurations are counted by
\begin{align}
z\frac{r}{s}\cdot [s^0]F(r,s) &= z\frac{r}{s}\cdot F(r,0).
\end{align}
\item This accounts for the possible pairs of walks without the interaction
parameters. We can now incorporate the interaction parameters. In order to
do this, we have to add in all walks we want to mark with $a$ and subtract the
non-weighted version of those exact same walks from the model
(Figure~\ref{int_terms} --- left). In order for the bottom walk to touch the
wall, it must be at height $1$ initially and then step down (with no restriction
on the top walk). Hence we get the term
\begin{align}
% za(1+s)\cdot [r^1]F(r,s) - z(1+s)\cdot [r^1]F(r,s) = 
z(a-1)(1+s)\cdot [r^1]F(r,s). 
\end{align}
\begin{figure}[h]
\begin{center}
 \includegraphics[width=14cm]{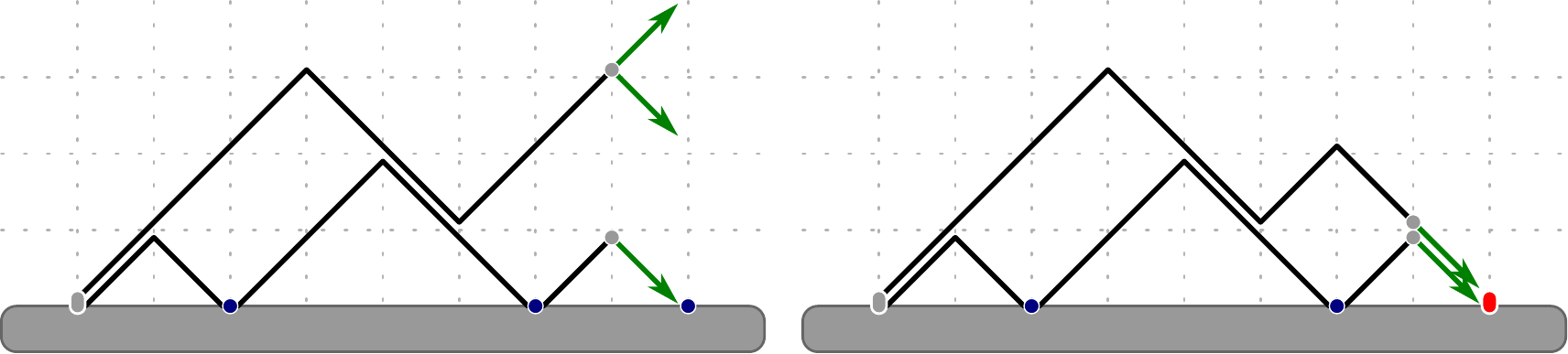}
\end{center}
\caption{Configurations that lead to single (left) and double (right)
interaction terms.}
\label{int_terms}
\end{figure}
\item A similar method can be used to incorporate $d$ into the model
(Figure~\ref{int_terms} --- right). One step before both walks touch the wall
they will both be at height $1$. All such walks have already been accounted for
when incorporating $a$ into the model and so must be replaced. This results in
\begin{align}
z(d-a)\cdot [r^1s^0]F(r,s).
\end{align}
\end{itemize}
The functions $[r^1]F(r,s)$ and $[r^1s^0]F(r,s)$ can be simplified in terms of
$F(0,0)$ and $F(0,s)$. By extracting the coefficient of $r^0s^0$ in the
functional equation, we obtain
\begin{align}
F(0,0) &= 1 + zd \cdot [r^1s^0]F(r,s).
\end{align}
At a combinatorial level, this states that a pair of walks that end at the wall
is either a trivial configuration or obtained by appending a pair of SE steps
to the end of a pair of walks that end at height $1$. Similarly, we can extract
the coefficient of $r^0$ in the functional equation to obtain
\begin{align}
F(0,s) = 1 + za(1+s) \cdot [r^1]F(r,s)  +z(d-a) \cdot [r^1s^0]F(r,s).
\end{align}
This has a similar combinatorial interpretation to the previous case. These
equations can then be combined to simplify the functional equation
\begin{multline}
\label{eqn full ad}
\left(1 - z\left[r + \frac{1}{r} + \frac{s}{r}+\frac{r}{s}\right]\right)
\cdot F(r,s) = \\
\frac{1}{d} +\left(1 - \frac{1}{a} - \frac{zs}{r} -
\frac{z}{r}\right)\cdot F(0,s) - \frac{zr}{s}\cdot F(r,0) + \left(\frac{1}{a} -
\frac{1}{d}\right) \cdot F(0,0).
\end{multline}
We will use the above form of the equation in what follows. The
polynomial coefficient on the left hand side is called the \emph{kernel} $K(r,s;z)
\equiv K(r,s)$, and its symmetries play a key role in the solution:
\begin{equation}
K(r,s) =\left[1 - z\left(r + \frac{1}{r} + \frac{s}{r}+\frac{r}{s}\right)\right].
\end{equation}

\section{Solution of the functional equations}
\label{sec:solfuneqn}
In what follows we use the obstinate kernel method. The discussion below is
self-contained, but we refer the reader to the paper of Bousquet-M\'elou and
Mishna \cite{bousquet2010a-a} for a general description of this technique.

\subsection{Solution of the functional equations when $a=1$}

When $a=1$, the functional equation~\Ref{eqn full ad} simplifies to 
% \begin{align}
% K(r,s) \cdot F(r,s) = \frac{1}{d} +\left(1 - \frac{1}{a} - \frac{zs}{r} - \frac{z}{r}\right)\cdot F(0,s) - \frac{zr}{s} F(r,0) + \left(\frac{1}{a} - \frac{1}{d}\right) F(0,0)
% \end{align}
% simplifies to 
\begin{align}
K(r,s) \cdot rsd F(r,s) = rs - zsd\left(1+s \right)\cdot F(0,s) - zr^2 d F(r,0) + rs\left(d-1\right) F(0,0).
\end{align}
We use the kernel method which exploits the symmetries of the kernel to remove
boundary terms (ie the functions $F(r,0)$ and $F(0,s)$) in the above equation.
The kernel is symmetric under the following two transformations:
\begin{align}
(r,s) &\mapsto \left(r,\frac{r^2}{s} \right), &
(r,s) &\mapsto \left(\frac{s}{r},s \right).
\end{align}
These transformations generate a family of 8 symmetries (sometimes referred to as
the `group of the walk' --- see \cite{bousquet2010a-a})
\begin{align}
  (r,s), 
  \left(r,\frac{r^2}{s} \right), 
  \left(\frac{s}{r},\frac{s}{r^2} \right),
  \left(\frac{r}{s},\frac{1}{s} \right),
  \left(\frac{1}{r},\frac{1}{s} \right),
  \left(\frac{1}{r},\frac{s}{r^2} \right),
  \left(\frac{r}{s},\frac{r^2}{s} \right), \text{ and }
  \left(\frac{s}{r}, s \right).
\end{align}
We make use of $4$ of these transformations --- those which only involve
positive powers of~$r$. To be precise,
\begin{subequations}
\footnotesize
\begin{align}
\label{eqn symm1}
K(r,s) \cdot rsd F(r,s) 
&= rs - zsd\left(1+s \right)\cdot F(0,s) - zr^2 d \cdot F(r,0) +
rs\left(d-1\right)\cdot F(0,0);\\
\label{eqn symm2}
K\left(r,\frac{r^2}{s} \right) \cdot \frac{dr^3}{s} F\left(r,\frac{r^2}{s}
\right) 
&= \frac{r^3}{s} - \frac{zdr^2}{s}\left(1+\frac{r^2}{s} \right)\cdot
F\left(0,\frac{r^2}{s}\right) - zr^2 d\cdot F(r,0) +
\frac{r^3}{s}\left(d-1\right) \cdot F(0,0);\\
\label{eqn symm3}
K\left(\frac{r}{s},\frac{r^2}{s} \right) \cdot \frac{dr^3}{s^2}
F\left(\frac{r}{s},\frac{r^2}{s} \right) 
&= \frac{r^3}{s^2} - \frac{zdr^2}{s}\left(1+\frac{r^2}{s} \right)\cdot
F\left(0,\frac{r^2}{s}\right) - \frac{zr^2 d}{s^2}\cdot  F(\frac{r}{s},0) +
\frac{r^3}{s^2}\left(d-1\right) \cdot F(0,0);\\
\label{eqn symm4}
K\left(\frac{r}{s},\frac{1}{s}\right) \cdot \frac{dr}{s^2}
F\left(\frac{r}{s},\frac{1}{s}\right) 
&= \frac{r}{s^2} - \frac{zd}{s}\left(1+\frac{1}{s} \right)\cdot
F\left(0,\frac{1}{s}\right) - \frac{zdr^2}{s^2}\cdot 
F\left(\frac{r}{s},0\right) +
\frac{r(d-1)}{s^2} \cdot F(0,0).
\end{align}
\end{subequations}
All of these transformations were chosen so that the kernel
remains unchanged, and so that the substitution only involves positive powers of
$r$. We can then eliminate the boundary terms by taking an alternating sum
of the above equations:
\begin{equation*}
\text{Eqn\Ref{eqn symm1}} - \text{Eqn\Ref{eqn symm2}}
+ \text{Eqn\Ref{eqn symm3}} - \text{Eqn\Ref{eqn symm4}}.
\end{equation*}
(In the case where $a \neq 1$, a similar method holds, except that then
we must multiply some of the equations by non-trivial coefficients to eliminate
boundary terms.) After simplification we obtain
\begin{multline}
K(r,s)\cdot\left(\text{linear combination of } F \right) = \\
\frac{r(s-1)(s^2+s+1 - r^2)}{s^2}\left(1 + (d-1) F(0,0)\right) \\
- zd(1+s)s F(0,s) + \frac{zd(1+s)}{s^2}F\left(0,\frac{1}{s}\right).
\end{multline}
We can now remove the left-hand side of the equation by choosing a value of $r$
that sets the kernel to zero --- provided all the $F$'s on the left-hand
side remain convergent. The kernel has two roots and we choose the one which
gives a positive term power series expansion in $z$ with Laurent polynomial
coefficients in $s$:
\begin{align}
 \hat{r}(s;z) \equiv \hat{r} &= 
\frac{s\left(1- \sqrt{1-4\frac{(1+s)^2z^2}{s} } \right)}{2(1+s)z} 
  = \sum_{n \geq 0} C_n \frac{(1+s)^{2n+1} z^{2n+1} }{s^n},
\end{align}
where $C_n = \frac{1}{n+1}\binom{2n}{n}$ is a Catalan number. This is chosen
so that $K(\hat{r},s) =0$, and so that all the various
substitutions are convergent. More precisely, since $\hat{r}=O(z)$, the
functions $F(\hat{r},s), F(\hat{r},\hat{r}^2/s), F(\hat{r}/s, \hat{r^2}/s)$ and
$F(\hat{r}/s,1/s)$ are all formally convergent power series in $z$ with Laurent
polynomial coefficients in $s$. 

We are not able to use the other root of the kernel (with respect to $r$) since
it is $O(z^{-1})$. If we were to substitute this into the functional equation,
then $F(r,s), F(r,r^2/s), F(r/s,r^2/s)$ and $F(r/s,1/s)$ would not converge
within the ring of formal power series. This follows since the coefficient of
$z^n$ in $F(r,s)$ has degree $n$ in $r$ and so substituting $r \mapsto
O(z^{-1})$ will map terms in this polynomial to all powers of $z$ including the
constant term.

When we make the substitution $r \mapsto \hat{r}$ we can rewrite the
coefficients of the right-hand side so as to not explicitly involve $z$ ---
since now $z = \left( \hat{r}+1/\hat{r} + \hat{r}/s + s/\hat{r} \right)^{-1}$.
\begin{align}
0 &= \frac{\hat{r}(s-1)(s^2+s+1 - \hat{r}^2)}{s^2}
\left(1 + (d-1) F(0,0)\right) -
\frac{d \hat{r} s^2}{s+\hat{r}^2} \,F(0,s) +
\frac{d \hat{r} }{(s+\hat{r}^2)s} \,F\left(0,\frac{1}{s}\right).
\end{align}

Because we are primarily interested in $F(0,0)$ --- the generating function
of pairs of walks that start and end on the wall --- it is convenient to rewrite
the equation so that there are no powers of $s$ or $\hat{r}$ in the
denominator of the coefficients and so that the coefficients of $F(0,s)$ and
$F(0,1/s)$ are independent of $\hat{r}$. 
\begin{align}
\label{eqn take s1}
0 &=d s^4 F(0,s) -  d s F\left(0,\frac{1}{s}\right) 
- (s-1)(s^2+s+1 -\hat{r}^2)(s+\hat{r}^2) \left(1 + (d-1) F(0,0)\right) .
\end{align}

Consider the coefficient of $s^1$ in the above equation, or rather by
dividing the equation by $s$ consider the \emph{constant term}, that is the
coefficient of $s^0$ in the equation. This leads us to calculate $F(0,0)$
effectively as a constant term in the variable $s$. For ease of calculation and
display we will continue with calculating the coefficient of $s^1$ in
equation~(\ref{eqn take s1}).

Since $F(0,s)$ is a power series in $z$ with polynomial coefficients in $s$, the
term $d s^4 F(0,s)$ does not contain any coefficients of $s^1$. Similarly,
$F\left(0,\frac{1}{s}\right)$ is a power series in $z$ with polynomial
coefficients in $s^{-1}$, so the term $dsF\left(0,\frac{1}{s}\right)$
contributes only $dF(0,0)$. For the remaining term, we consider the coefficient
$[s^1] (s-1)(s^2+s+1 - \hat{r}^2)(s+\hat{r}^2)$. Expanding the expression and
then collecting the exponents of $\hat{r}$ gives:
\begin{align}
 (s-1) \hat{r}^4+ (1-s+s^2-s^3) \hat{r}^2+ s(1-s)(s^2+s+1).
\end{align}
We need to consider the expansion of $\hat{r}^2$ and $\hat{r}^4$. Lagrange
inversion \cite{flajolet2009a-a} gives: 
\begin{subequations}
\begin{align}
\hat{r}(s;z) &= \sum_{n=0}^{\infty} \frac{C_n (1+s)^{2n+1}}{s^n} z^{2n+1},\\
\hat{r}(s;z)^2 &= \sum_{n=0}^{\infty} \frac{C_{n+1} (1+s)^{2n+2}}{s^n}
z^{2n+2},\\
\hat{r}(s;z)^4 &= \sum_{n=0}^{\infty}
\frac{4}{2n+4}\binom{2n+4}{n}\frac{(1+s)^{2n+4}}{s^n} z^{2n+4},
\intertext{and, more generally,}
\hat{r}(s;z)^k &= \sum_{n=0}^{\infty}
\frac{k}{2n+k}\binom{2n+k}{n}\frac{(1+s)^{2n+k}}{s^n} z^{2n+k}.
\label{eqn power of r}
\end{align}
\end{subequations}

Computing the coefficient of a particular power of $s$ in $\hat{r}^2$ or
$\hat{r}^4$ reduces to finding the coefficient of powers of $s$ in $(1+s)^n
s^{-k}$ which are just binomial coefficients:
\begin{subequations}
\begin{align}
[s^1] (s-1) \hat{r}^4 &= \sum_{n=0}^{\infty} -\frac{6(n-1)n}{(n+2)^2}C_n^2
z^{2n+2};\\
[s^1] (1-s+s^2-s^3) \hat{r}^2 &=\sum_{n=0}^{\infty}
\frac{6(n^2+1)}{(n+2)(n+3)} C_n^2 z^{2n+2};\\
[s^1] s(1-s)(s^2+s+1) &= 1 ;\\
[s^1] (s-1)(s^2+s+1 - \hat{r}^2)(s+\hat{r}^2) &= 1 + \sum_{n=0}^{\infty}
\frac{12(2n+1)}{(n+2)^2(n+3)} C_n^2 z^{2n+2}.
\end{align}
\end{subequations}
Hence extracting the coefficient of $s^1$ in equation~\Ref{eqn take s1} gives
\begin{align}
0&= - \left(1 + \sum_{n=0}^{\infty} \frac{12(2n+1)}{(n+2)^2(n+3)} C_n^2
z^{2n+2}\right)\cdot\left(1 + (d-1) F(0,0)\right) - d\cdot F(0,0).
\end{align}
Solving the above when $d=1$ gives
\begin{align}
\label{eqn g11}
G(1,1;z) &= 1 + \sum_{n=0}^{\infty} \frac{12(2n+1)}{(n+2)^2(n+3)} C_n^2 z^{2n+2},
\end{align}
and hence for general $d$ we have
\begin{align}
\label{eqn g1d}
F(0,0) = G(1,d;z) &= \frac{G(1,1;z)}{d + (1-d)G(1,1;z)}.
\end{align}
In Section~\ref{sec:altsol} we will see that the algebraic structure of this
solution that gives $ G(1,d;z)$ in terms of $G(1,1;z)$ arises naturally from a
combinatorial construction. Moreover, this structure extends to the $a\neq 1$
case.

\subsection{Solution of the functional equation when $a\neq1$}

The general $a,d$ case can be solved by the method applied above, however, it is
sufficient to study the case $d=a$ which can be resolved more cleanly. As
mentioned above the algebraic structure that allows $ G(1,d;z)$ to be expressed
in terms of $G(1,1;z)$ extends to give $ G(a,d;z)$ in terms of $G(a,a;z)$. We
shall see that explicitly in Section~\ref{sec:altsol}. When $d=a$ the functional
equation~\Ref{eqn full ad} simplifies to
\begin{align}
K(r,s)\cdot F(r,s)a^2rs &= (ar-r-za-zas)as\cdot F(0,s)-zr^2a^2\cdot F(r,0)+ars .
\end{align}
The symmetries we used above can be reused to remove boundary terms. As above we
take an alternating sum of transformed equations, but now we must multiply some
of the equations by a non-trivial factor chosen to eliminate boundary terms. The
left-hand side becomes
\begin{align}
LHS & = a^2 r K(r,s) \left(s F(r,s) - \frac{r^2}{s}
F\left(r,\frac{r^2}{s}\right) +
\frac{L r^2}{s^2}F\left(\frac{r}{s},\frac{r^2}{s}
\right) - \frac{L}{s^2}
F\left(\frac{r}{s},\frac{1}{s}\right) \right),
\end{align}
where 
\begin{align}
L  &= \frac{zas-ars+rs+zar^2}{zas-ar+r+zar^2}.
\end{align}
The right-hand side simplifies to
\begin{multline}
\label{eqn aneq1 kern}
RHS
=as^2(1+s-a)F(0,s)+a(1+s-as)F\left(0,1/s\right)\\
-\frac{(r^2+s)a(s-1)(ar^2+as-2s-s^2-1)}{ar^2-r^2-s}.
\end{multline}
Again, we have attempted to massage the functional equation into a form in which
the coefficients of $F(0,s)$ and $F(0,1/s)$ are independent of $r$.
Unfortunately, we cannot completely clear the denominator of the above
functional equation, and we found it simplest to work with the above expression.

Following the method used in the $a=1$ case, we can eliminate the
left-hand side further by choosing a value of $r$ that sets the kernel to $0$.
We choose the root which gives a positive term power series expansion in $z$
with Laurent polynomial coefficients in $s$. Recall that $\hat{r}$ is given by
\begin{align}
\hat{r}(s;z) \equiv \hat{r} 
&= \sum_{n\geq0} C_n \frac{(1+s)^{2n+1} z^{2n+1}}{s^n}.
\end{align}

Substituting $r \mapsto \hat{r}$ eliminates the left-hand side of the
functional equation and we again consider the coefficient of $s^1$ in the
resulting right-hand side. Again, this can be converted into a \emph{constant
term} expression for our generating function. The term $as^2(1+s-a)F(0,s)$
does not contribute to $s^1$. The term $a(1+s-as)F\left(0,1/s\right)$
contributes $a(1-a)F\left(0,0\right)$. For the remaining term, we consider the
expansion of the expression as a series
in $a$. The coefficient of $a^1$ is %as opposed to a^2
\begin{align}
[a^1]%a^2
\frac{a(\hat{r}^2+s)(s-1)(a\hat{r}^2+as-2s-s^2-1)}{a\hat{r}^2-\hat{r}^2-s } 
&= (1-s)(1+s)^2,
\end{align}
and hence
\begin{align}
[a^1s^1]%a^2
\frac{a(\hat{r}^2+s)(s-1)(a\hat{r}^2+as-2s-s^2-1)}{a\hat{r}^2-\hat{r}^2-s } 
&= -1.
\end{align}
Higher powers of $a$ are (for $k \geq 1$)
\begin{multline}
\label{eqn:twoterms}
[a^{k+1}]
\frac{a(\hat{r}^2+s)(s-1)(a\hat{r}^2+as-2s-s^2-1)}{a\hat{r}^2-\hat{r}^2-s} 
\\
\begin{split}
= & \frac{(s-1)(\hat{r}-1)(\hat{r}+1)(\hat{r}-s)(\hat{r}+s)}{\hat{r}^2}\cdot
\left(\frac{\hat{r}^2}{s+\hat{r}^2}\right)^{k}\\
= & \left( \frac{(s-1)s^2}{(s+1)^2z^2}-(s+1)^2(s-1)\right)\cdot
\left(\frac{\hat{r}^2}{s+\hat{r}^2}\right)^{k}\\
= & \left( \frac{(s-1)s^2}{(s+1)^2z^2}\right)\cdot
\left(\frac{\hat{r}^2}{s+\hat{r}^2}\right)^{k}
-\left((s+1)^2(s-1)\right)\cdot\left(\frac{\hat{r}^2}{s+\hat{r}^2}\right)^{k}.
\end{split}
\end{multline}
To extract the coefficient of
$s^1$, we need to consider the expansion of $\left(\frac{\hat{r}^2}{s+
\hat{r}^2}\right)^{k}$ in $z$. This exponential term simplifies, and we
can use equation~\Ref{eqn power of r} to obtain
\begin{align}
\left(\frac{\hat{r}^2}{s+ \hat{r}^2}\right)^{k} 
= z^k \left( \frac{1+s}{s} \right)^k \hat{r}^k
&=
\sum_{p\geq0}\frac{k}{2p+k}\binom{2p+k}{p}\frac{(s+1)^{2p+2k}}{s^{p+k}}z^{2p+2k}.
\end{align}

We will expand the two terms in \Ref{eqn:twoterms} individually. For the first
term, we get
\begin{multline}
[s^1]\frac{(s-1)s^2}{(s+1)^2z^2}\cdot
\left(\frac{\hat{r}^2}{s+\hat{r}^2}\right)^{k} \\
= \sum_{p\geq0}\frac{k}{k+2p}\binom{k+2p}{p}\left(\binom{2k-2+2p}{k+p-1} -
\binom{2k-2+2p}{k+p-2}\right)z^{2k-2+2p}.
\end{multline}
We can extract the coefficient of $z^{2n}$ from the above equation by making the
substitution $n = k-1+p$. We obtain
\begin{align}
[z^{2n}s^1]\frac{(s-1)s^2}{(s+1)^2z^2}\cdot
\left(\frac{\hat{r}^2}{s+\hat{r}^2}\right)^{k} 
&= \frac{k}{2n-k+2}\binom{2n-k+2}{n+1}\left[\binom{2n}{n} -
\binom{2n}{n-1}\right].
\end{align}
Therefore
\begin{align}
[s^1]\frac{(s-1)s^2}{(s+1)^2z^2}\cdot
\left(\frac{\hat{r}^2}{s+\hat{r}^2}\right)^{k} 
&= \sum_{n\geq k-1} \frac{k}{2n-k+2}\binom{2n-k+2}{n+1}\left[\binom{2n}{n} -
\binom{2n}{n-1}\right] z^{2n}.
\end{align}

Following a similar argument for the second term in equation~\Ref{eqn:twoterms},
we have
\begin{multline}
[s^1](s+1)^2(s-1)\cdot \left(\frac{\hat{r}^2}{s+\hat{r}^2}\right)^{k}\\
= \sum_{p\geq0}\frac{k}{k+2p}\binom{k+2p}{p}\left(\binom{2k+2p+2}{k+p}-\binom{
2k+2p+2}{k+p-1}\right)z^{2k+2p}.
\end{multline}
Making the substitution $n = k+p$, we get
\begin{align}
[z^{2n}s^1](s+1)^2(s-1)\cdot \left(\frac{\hat{r}^2}{s+\hat{r}^2}\right)^{k}&= 
\frac{k}{2n-k}\binom{2n-k}{n}\left[\binom{2n+2}{n+1}-\binom{2n+2}{n}\right].
\end{align}
We can then substitute the summation over $p$ with a summation over $n$.
\begin{align}
[s^1](s+1)^2(s-1)\cdot \left(\frac{\hat{r}^2}{s+\hat{r}^2}\right)^{k}&= \sum_{n\geq k}
\frac{k}{2n-k}\binom{2n-k}{n}\left[\binom{2n+2}{n+1}-\binom{2n+2}{n}\right] z^{2n}.
\end{align}
When $n = k-1$ in the above equation, the summand reduces to $0$ when $k>2$. So
it is possible to adjust the range of the summation by adjusting for the $k =
1,2$ cases separately. In those cases, the combined correction terms are $a^2$
and $-4a^3z^2$ respectively. Thus, we can rewrite
\begin{multline}
[s^1](s+1)^2(s-1)\cdot \left(\frac{\hat{r}^2}{s+\hat{r}^2}\right)^{k}\\= \sum_{n\geq k-1}
\frac{k}{2n-k}\binom{2n-k}{n}\left[\binom{2n+2}{n+1}-\binom{2n+2}{n}\right]
z^{2n},
\end{multline}
with known correction terms for $k=1,2$. Combining these summands, we get that for $n\geq k-1$:
\begin{multline}
\frac{k}{2n-k+2}\binom{2n-k+2}{n+1}\left[\binom{2n}{n} - \binom{2n}{n-1}\right] \\- \frac{k}{2n-k}\binom{2n-k}{n}\left[\binom{2n+2}{n+1}-\binom{2n+2}{n}\right] \\= \frac{k(k+1)(2+4n-kn-2k)}{(k-1-n)(n+1)^2(k-2n)(n+2)}\binom{2n-k}{n}\binom{2n}{n}.
\end{multline}
Thus taking the coefficient of $s^1$ when $r = \hat{r}$ in equation~\Ref{eqn
aneq1 kern} and accounting for the correction terms, we get
\begin{multline}
% \label{eqn:qnks}
% [s^1]0 &=[s^1]as^2(1+s-a)F(0,s)+[s^1]a(1+s-as)F\left(0,1/s\right)-[s^1]\frac{a(r^2+s)a(s-1)(ar^2a+as-2s-s^2-1)}{ar^2-r^2-s}\\
0 = a(a-1) F(0,0)  - a + a^2 - 4z^2a^3\\
- \sum_{k\geq1} a^{k+1} \sum_{n\geq k-1}
\frac{k(k+1)(2+4n-kn-2k)}{(k-1-n)(n+1)^2(-2n+k)(n+2)}\binom{2n-k}{n}\binom{2n}{n
} z^{2n}.
\end{multline}
We exchange the order of summation to give
\begin{multline}
\label{eqn:qnks}
0 = a(a-1) F(0,0)  - a + a^2 - 4z^2a^3\\
- \sum_{n\geq0} z^{2n} \sum_{ k=1}^{n+1}
\frac{k(k+1)(2+4n-kn-2k)}{(k-1-n)(n+1)^2(-2n+k)(n+2)}\binom{2n-k}{n}\binom{2n}{n
} a^{k+1}.
\end{multline}

The extraction of the coefficient $[a^kz^{2n}]F(0,0)$ requires rearranging the
$a(a-1)$ coefficient in front of $F(0,0)$. We express the above equation
as:
\begin{equation}
0 = a(a-1) F(0,0) - \sum_{n\geq0} z^{2n} \sum_{ k'=1}^{n+1} Q_{n,k'} a^{k'+1}
\end{equation}
for some integers $Q_{n,k'}$. This can be rearranged to give:
\begin{align}
F(0,0) &= - \left(\sum_{n\geq0} z^{2n} \sum_{ k'=1}^{n+1} Q_{n,k'} a^{k'}\right)\cdot \frac{1}{1-a}\\
&= - \left(\sum_{n\geq0} z^{2n} \sum_{ k'=1}^{n+1} Q_{n,k'} a^{k'}\right)\cdot \left(\sum_{k''\geq0}a^{k''}\right).
\end{align}
The coefficient of $a^k$ from the above is summation of all contributions from
$k'$ and $k''$ such that $k'+k'' =k$. Thus:
\begin{equation}
F(0,0) = \sum_{n\geq0} z^{2n} \sum_{ k=1}^{n+1} a^k \sum_{k'=1}^{k}Q_{n,k'}.
%\sum_{n,k\geq 0} \left(\sum_{k'=0}^{k} Q_{n,k'} \right) a^{k}z^{2n}.
\end{equation}
In other words, extracting the coefficient $[a^kz^{2n}]F(0,0)$ requires a
summation of a finite number of the $Q_{n,k'}$ terms which is obtained
from~\Ref{eqn:qnks}.
\begin{align}
[a^kz^{2n}]F(0,0) &= \sum_{k' = 0}^k \frac{k'(k'+1)(2+4n-k'n-2k')}{(k'-1-n)(n+1)^2(-2n+k')(n+2)}\binom{2n-k'}{n}\binom{2n}{n}\\
&= \frac{k(k+1)(k+2)}{(2n-k)(n+1)^2(n+2)}\binom{2n-k}{n}\binom{2n}{n}.
\end{align}
We finally obtain
\begin{align}
\label{eqn_ct_gaa}
F(0,0) = G(a,a)  &=
\sum_{n\geq0}
z^{2n}
\sum_{k=0}^n
a^k \frac{k(k+1)(k+2)}{(n+1)^2(n+2)(2n-k)}\binom{2n}{n}\binom{2n-k}{n}.
\end{align}

This agrees with results due to Brak \emph{et\ al.}\  \cite{brak1998c-:a,
owczarek2001c-:a} for a closely related model  obtained using very different
method --- we discuss this more fully in the section~\ref{sec:altsol}.

\section{Alternate solution}
\label{sec:altsol}
An alternate technique for finding the generating function relies on factoring
the pairs of walks at each double-visit. First, let us define
\begin{align}
 G(a,d) \equiv G(a,d;z) &= F(0,0;a,d;z).
\end{align}
We will frequently hide the $z$ dependence for convenience. 
Breaking up our configurations into pieces between double visits gives
\begin{align}
\label{eqn gad pa}
 G(a,d;z) &= \frac{1}{1-dP(a;z)},
\end{align}
where $P(a;z)$ is the generating function of so-called primitive factors. This
is quite analogous to the classical factorisation of a single Dyck path. These
primitive factors are pairs of friendly Dyck-paths which contain no
double-visits to the surface other than their first and last vertices.
Rearranging this expression 
gives
\begin{align}
\label{eqn pa gad gaa}
 P(a;z) &= \frac{G(a,d;z)-1}{dG(a,d;z)} = \frac{G(a,a;z)-1}{aG(a,a;z)}.
\end{align}
This last expression allows us to calculate $P(a;z)$ from a known expression
for $G(a,a;z)$ --- such as that given in the previous section. Alternatively,
one could use results from previous work by Brak \emph{et\ al.}\  \cite{brak1998c-:a,
owczarek2001c-:a}. In those works, the authors considered a vesicle model which
corresponds exactly to the case $d=a$ --- their vicious walk model can be
transformed into the friendly walk model considered here, by moving the upper
vesicle boundary down by 2 units.

Brak \emph{et\ al.} use the Lindstr\"om-Gessel-Viennot lemma
\cite{gessel1985a-a, lindstrom1973a-a} to express the partition function of the
pair of walks in terms of the partition function of a single walk. Namely,
\begin{align}
\label{eqn Gaa GV}
 [z^n] G(a,a) &= \frac{S_{2n}(1) S_{2n+4}(a) - S_{2n+2}(1)
S_{2n+2}(a) }{a^2},
\end{align}
where $S_{2n}(a)$ is the partition function of a single Dyck path of length $2n$
above a wall, and $a$ is conjugate to the number of visits, ie
\begin{align}
 S_{2n}(a) &= [z^{2n}] \, 2 \left( 2 - a + a\sqrt{1-4 z^2} \right)^{-1} \\
  &= \sum_{k=0}^n \frac{2k+1}{n+k+1} \binom{2n}{n-k} (a-1)^k
  = \sum_{k=1}^n \frac{k}{2n-k} \binom{2n-k}{n} a^k,
\end{align}
where this last formula is taken from \cite{brak1998c-:a}. When $a=1$ we recover
the well-known Catalan number result, and a well-known central binomial result
when $a=2$:
\begin{align}
  S_{2n}(1) &= C_n = \frac{1}{n+1} \binom{2n}{n},
  &
  S_{2n}(2) &= \binom{2n}{n}.
\end{align}
In light of these simple expressions one can write $G(a,a;z)$ as double sum of
products of binomials. Using equation~\Ref{eqn pa gad gaa} we write $G(a,d;z)$
in terms of $G(a,a;z)$
\begin{align}
\label{eqn Gad Gaa}
 G(a,d) &= \frac{a G(a,a)}{d + (a-d)G(a,a)},
%   = \frac{G(a,a)}{1 - \frac{d}{a}\left(G(a,a)-1 \right)}
\intertext{where}
 G(a,a) &= a^{-2} \sum_{n=0}^{\infty}\left[
  C_n S_{2n+4}(a) - C_{n+1} S_{2n+2}(a)
\right] z^{2n},
\end{align}
which simplifies to the expression in equation~\Ref{eqn_ct_gaa} found in  the
previous section.

\subsection{Solutions at $a=0,1$ and $2$}
\label{sec sol a012}
Since the partition function $S_{2n}(a)$ takes simple values at $a=0,1,2$, we
have
\begin{align}
 G(1,1;z) &= \sum_{n=0}^\infty \left[ C_n C_{n+2} - C_{n+1}^2 \right] z^{2n},
\nonumber\\
\label{eqn G11}
  &= \sum_{n=0}^\infty \frac{12(2n+1)}{(n+1)^2(n+2)^2(n+3)}
\binom{2n}{n}^2 z^{2n},\\
\label{eqn G22}
 G(2,2;z) &= \sum_{n=0}^\infty C_n C_{n+1} z^{2n}, \\
 \intertext{and} \label{eqn G0}
 \lim_{a \to 0} \frac{G(a,a;z)-1}{a} 
% &= \sum_{n=1}^\infty \left[ C_{n-1} C_{n+1} - C_n^2 \right] z^{2n} 
% \nonumber\\
& = \sum_{n=1}^\infty 
\frac{12(2n-1)}{n^2(n+1)^2(n+2)}
\binom{2n-2}{n-1}^2 z^{2n} 
= z^2 G(1,1;z).
\end{align}

We can use these together with equation~\Ref{eqn Gad Gaa} to derive expressions
for $G(1,d)$ (that agrees with~\Ref{eqn g11} and~\Ref{eqn g1d}) and $G(2,d)$ by
simple substitutions. That is,
\begin{align}
\label{eqn G1d G2d}
 G(1,d) &= \frac{  \sum_{n=0}^\infty \frac{12(2n+1)}{(n+1)^2(n+2)^2(n+3)}
\binom{2n}{n}^2 z^{2n}}{d + (1-d) \sum_{n=0}^\infty \frac{12(2n+1)}{(n+1)^2(n+2)^2(n+3)}
\binom{2n}{n}^2 z^{2n}}
 \intertext{and}
 G(2,d) &= \frac{2 \sum_{n=0}^\infty C_n C_{n+1} z^{2n}}{d + (2-d)\sum_{n=0}^\infty C_n C_{n+1} z^{2n}}.
 \end{align}
A little further work also gives
\begin{align}
 G(0,d) &= \frac{1}{1-d z^2 G(1,1)} =  \frac{1}{1-d z^2 \sum_{n=0}^\infty \frac{12(2n+1)}{(n+1)^2(n+2)^2(n+3)}
\binom{2n}{n}^2 z^{2n}}.
\end{align}
This last expression can be derived combinatorially by noting that in the limit
$a \to 0$ single visits are forbidden. In this limit, the primitive pieces are
in bijection with all walks counted by $G(1,1)$; any primitive piece can be
transformed into a pair of walks counted by $G(1,1)$ by moving them 1 lattice
unit up and gluing edges at the start and end.

\section{Analysis of phase structure and transitions}
\label{sec:analpd}
\subsection{Phases}
We now turn to the phase diagram of the model which is dictated by the radius
of convergence of $G(a,d;z)$ as a power series in $z$. Denote the radius of
convergence by $z_c(a,d)$. Equation~\Ref{eqn gad pa} shows that the
singularities of $G(a,d;z)$ are those of $P(a;z)$ and the simple pole at
$1-dP(a;z) = 0$. Denote this latter singularity by $z_d(a,d)$.
Equation~\Ref{eqn pa gad gaa} shows that the singularities of $P(a;z)$ are
related to those of $G(a,a;z)$ which are known from \cite{brak1998c-:a,
owczarek2001c-:a}.

In particular, the radius of convergence of $G(a,a;z)$ is
\begin{align}
\label{eqn rhoa}
 \rho(a) = \text{r.o.c } G(a,a;z) &= 
\begin{cases}
 \frac{1}{4} = z_b & a \leq 2, \\
 \frac{\sqrt{a-1}}{2a} = z_a(a) & a > 2.
\end{cases}
\end{align}
For $a<2$, the thermodynamic phase is related to $z_b$ and is the desorbed phase
in which the walks drift away from the wall and the mean number of visits is
$O(1)$. When $a>2$, $z_a(a)$ dominates and the lower walk adsorbs onto the wall
and the number of visits is $O(n)$. At $a=2$, there is a second-order phase
transition and a jump discontinuity in the specific heat (the second derivative
of the free energy). In both of these phases, the upper walk drifts away from
the wall, and the number of doubly-visited vertices is $O(1)$.

In the full $a$-$d$ model there are 3 phases, two of which are described in the
previous paragraph. In the third phase, associated with the simple pole at
$z_d(a,d)$, we shall see that the number of doubly-visited vertices is $O(n)$.
In what follows, we name these three phases associated with $z_b, z_a$ and 
$z_d$, desorbed, $a$-rich and $d$-rich respectively.

\subsection{Desorbed to $a$-rich transition}
In \cite{brak1998c-:a}, it was shown that the asymptotic behaviour of the
singular part of $G(a,a;z)$ near its radius of convergence is given by
\begin{align}
\label{eqn gaa sing}
 G(a,a;z) & \sim 
\begin{cases}
 A_- (1-4z)^4 \log(1-4z) & a<2, \\
 A_0 (1-4z)^2 \log(1-4z) & a=2, \\
 A_+ \left(1-z/z_a(a) \right)^{1/2} & a>2,
\end{cases}
\end{align}
where $z_a(a)=\frac{\sqrt{a-1}}{2a}$. It is important to notice that for all
$a$, the singularities are convergent and therefore $G(a,a;z)$ is convergent on
its radius of convergence $\rho(a)$.

If we fix $d$ at some small value, and then increase $z$ from 0 towards
$\rho(a)$, then $P(a;z)$ increases from $0$ to $P(a;\rho(a))$. Since $d$ is
small, and $P(a;\rho(a))$ is finite, $1 - d P(a;\rho(a) >0$ and so the only
singularities of $G(a,d;z)$ will be those of $P(a;z)$ and so those of
$G(a,a;z)$.

Thus for small values of $d$ there is a phase transition on moving $a$ 
through $2$ which describes the transition from the desorbed phase to an
$a$-rich phase as occurs in \cite{brak1998c-:a}. This adsorption transition has
been well-studied previously and is unusual in that it has a jump
discontinuity in the second derivative of the free-energy rather than a
divergence.

\subsection{Desorbed to $d$-rich transition}
Let us restrict our attention to $a<2$ and consider the effect of increasing
$d$. The argument in the previous subsection breaks down as soon as $1-
dP(a;\rho(a))=0$. Call this value $d_c(a) = P(a;\rho(a))^{-1}$.

Fix $a<2$ and $d>d_c$ and consider increasing $z$ from $0$ towards $\rho(a)
= 1/4$. The function $P(a;z)$ is an increasing function of $z$ (since it is a
positive term power series) and so it increases towards $P(a;\rho(a))$. Since
$d>d_c$, $P(a;z)$ will reach the value $d^{-1}$ before it reaches
$P(a,\rho(a))$ and the simple pole will occur when $z=z_d$, where $z_d$ is the solution of
\begin{align}
  \label{eqn P dinv}
  P(a;z_d(a,d) ) &= d^{-1}
\end{align}
and $z_d<z_b = 1/4$ in this region.

Hence, for $a<2$, there is a phase transition where $z_d$ and $z_b$ coincide at
\begin{align}
\label{eqn dca}
 d_c(a) &= P(a;1/4)^{-1} = \frac{a G(a,a;1/4)}{G(a,a;1/4)-1}.	
\end{align}
In order to determine the density of the singly- and doubly-visited vertices in
the $d>d_c$ phase consider the partial derivatives of $z_d(a,d)$ with respect to $a$
and $d$. Since $z_d(a,d)$ is defined by $d P(a; z_d(a,d)) = 1$, the
derivatives of $z_d(a,d)$ with respect to $a$ and $d$ are non-zero and so there are positive
densities of both singly- and doubly-visited vertices.

Now let us turn to the order of this transition; this can be determined by
examining the behaviour of $P(a;z)$ close to $z=1/4$ which is determined by the
behaviour of $G(a,a;z)$ --- see equation~\Ref{eqn pa gad gaa}. Close to $z=1/4$
we can write
\begin{align}
 G(a,a;z) &= G_{analytic}(a;z) + G_{singular}(a;z),
\end{align}
where the behaviour of $G_{singular}$ is given by equation~\Ref{eqn gaa sing}.
Consider an expansion of $G_{analytic}$ about $z=1/4$
\begin{align}
 G_{analytic}(a;z) & \approx G(a,a;1/4) + c_1(1-4z) + \cdots
\end{align}
for some non-zero constant $c_1$. The linear correction dominates
the dominant singular term in $G_{singular}$. Expanding equation~\Ref{eqn P
dinv} about $z=1/4$ gives
\begin{align}
  \frac{1}{d_c} - \frac{1}{d} &=  P(a;1/4) - P(a;z_d), \nonumber\\
  \frac{d-d_c}{d_c^2} &\approx  p_1(1-4z_d)
\end{align}
for some non-zero constant $p_1$. Hence there is a linear relationship between
the location of the $d$-rich singularity, $z_d$, and the distance from the
phase boundary. Since the free-energy of the system is $-\log z_d$, this also
implies the free-energy in the $d$-rich phase changes linearly with $d$. On the
other hand in the desorbed phase where $z_b$ dominates, the free-energy is a
constant. From this we see that there is a jump discontinuity in the first
derivative of the free-energy and hence this is a first-order transition. Note
that the above argument will also work \emph{mutatis mutandis} at $a=2$.

We can observe at finite length a characteristic bimodal probability
distribution in the number of doubly-visited vertices --- see Figure~\ref{fig
double peak}.

\begin{figure}[h]
 \begin{center}
  \includegraphics[width=12cm]{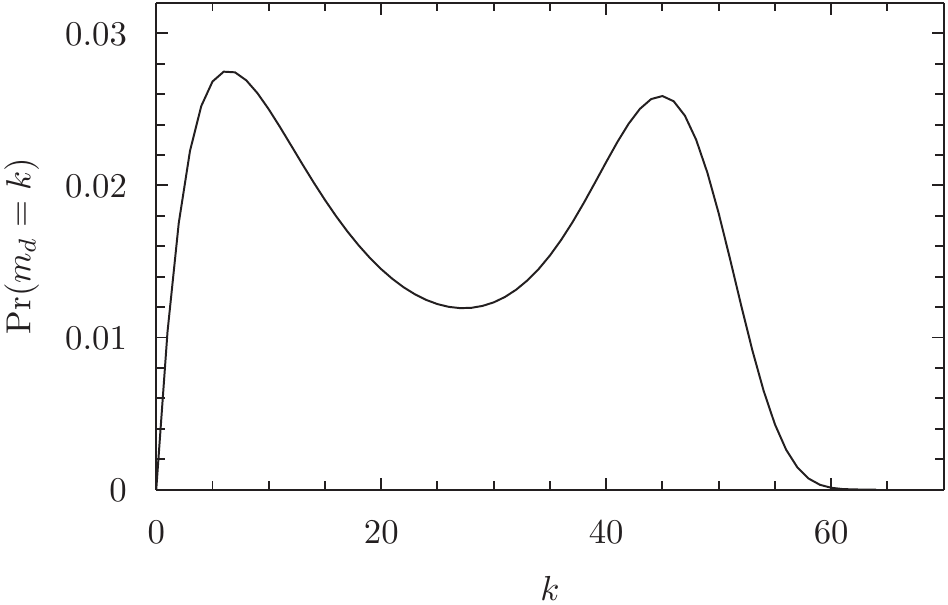}\\
 \end{center}
 \caption{A plot of the probability that a conformation of length 128 has $k$
doubly-visited sites at $a=1, d=10.3$. This value of $d$ corresponds to the
approximate location of the peak in the specific heat at this length.}
\label{fig double peak}
\end{figure}

\subsection{$a$-rich to $d$-rich transition}

The analysis of the previous section can be adapted to the case $a>2$ with some
important differences. The transition is driven by the singularities $z_a(a) =
\frac{\sqrt{a-1}}{2a}$ associated with single-visit adsorption, and the
singularity $z_d(a,d)$ associated with double-visit adsorption. Again,
$z_d(a,d)$ is the solution of equation~\Ref{eqn P dinv}. These two
singularities coincide when $d = d_c(a)$ given by
\begin{align}
  d_c(a) &= P\left(a;\frac{\sqrt{a-1}}{2a}\right)^{-1}
 = \frac{
a G\left(a,a;\frac{\sqrt{a-1}}{2a}\right)
}{
G\left(a,a;\frac{\sqrt{a-1}}{2a} \right)-1
}.
\end{align}

Turning to the order of this transition, we again decompose $G(a,a;z)$ into its
analytic and singular parts. Observe that close to $z_a(a)$, $G_{singular}$,
given by equation~\Ref{eqn gaa sing}, dominates the linear part of
$G_{analytic}$. Hence we deduce that 
\begin{align}
  d_c(a)-d & \approx p_2 \left(\frac{\sqrt{a-1}}{2a} - z_d(a,d) \right)^{1/2}, \\
  z_d(a,d) &\approx  \frac{\sqrt{a-1}}{2a} + p_3 \left(d - d_c(a) \right)^2
\end{align}
for some nonzero constants $p_2, p_3$. Therefore the free-energy has a jump
discontinuity in its second derivative on varying $d$ across the transition, and
this is a second-order phase transition. This is very similar to the
desorbed to $a$-rich transition.

\subsection{Phase diagram}
We have established that there are 3 thermodynamic phases; desorbed, $a$-rich
and $d$-rich. We remind the reader that $m_a(\varphi)$ and $m_d(\varphi)$
denote the number of single and double visits of $\varphi$.

\begin{figure}[h]
\begin{center}
 \includegraphics[height=6cm]{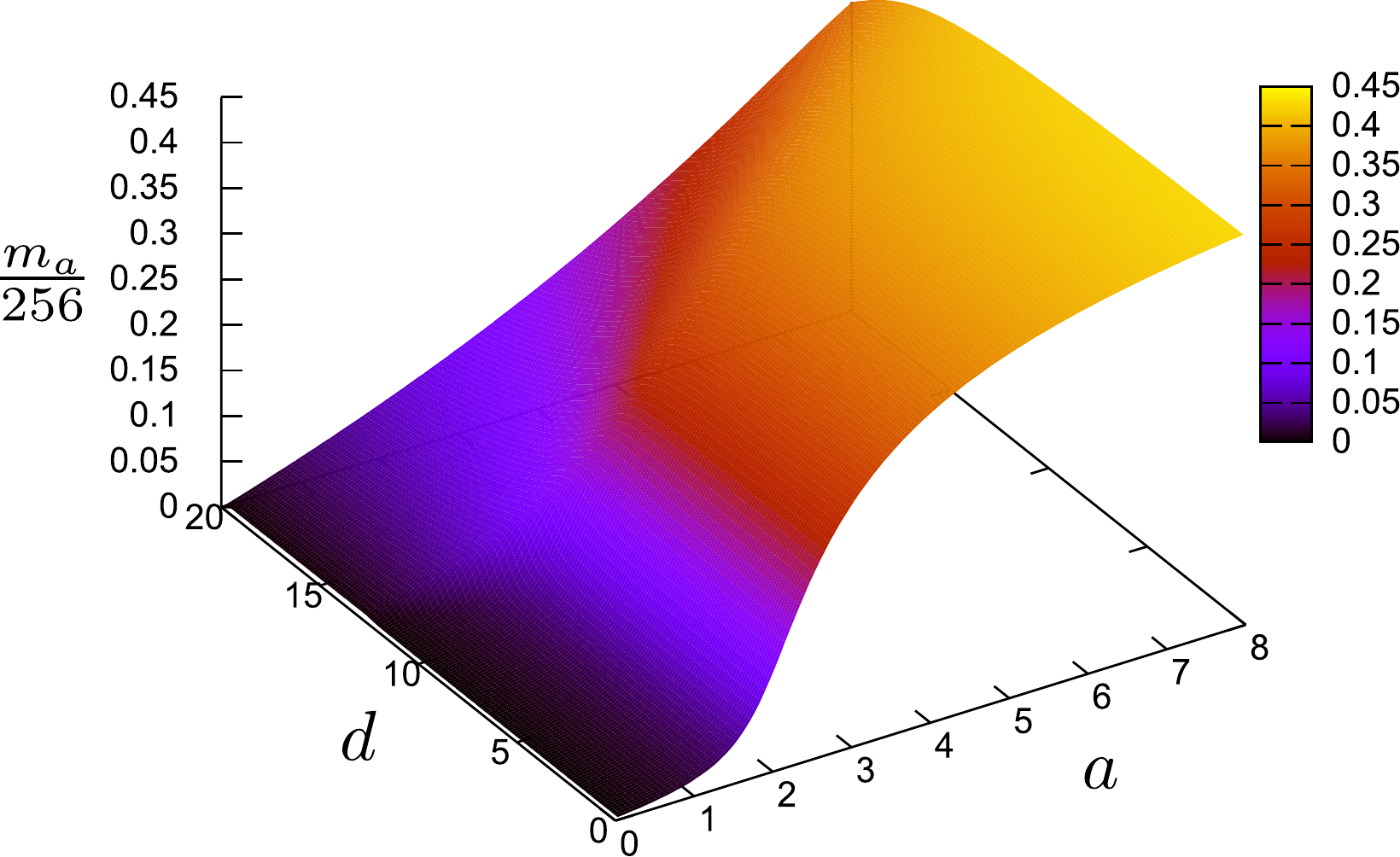}
\end{center}
\caption{A plot of the density of $a$ visits calculated at length $n=256$. This
highlights the region where it tends to a non-zero constant and corresponds
well to the regions where $z_a$ and $z_d$ dominate. Note that for fixed $a$ and
increasing, large $d$ we expect that the density of $a$ visits decreases
though remains positive.}
\label{ma ord parameter}
\end{figure}

\begin{figure}[h]
\begin{center}
 \includegraphics[height=6cm]{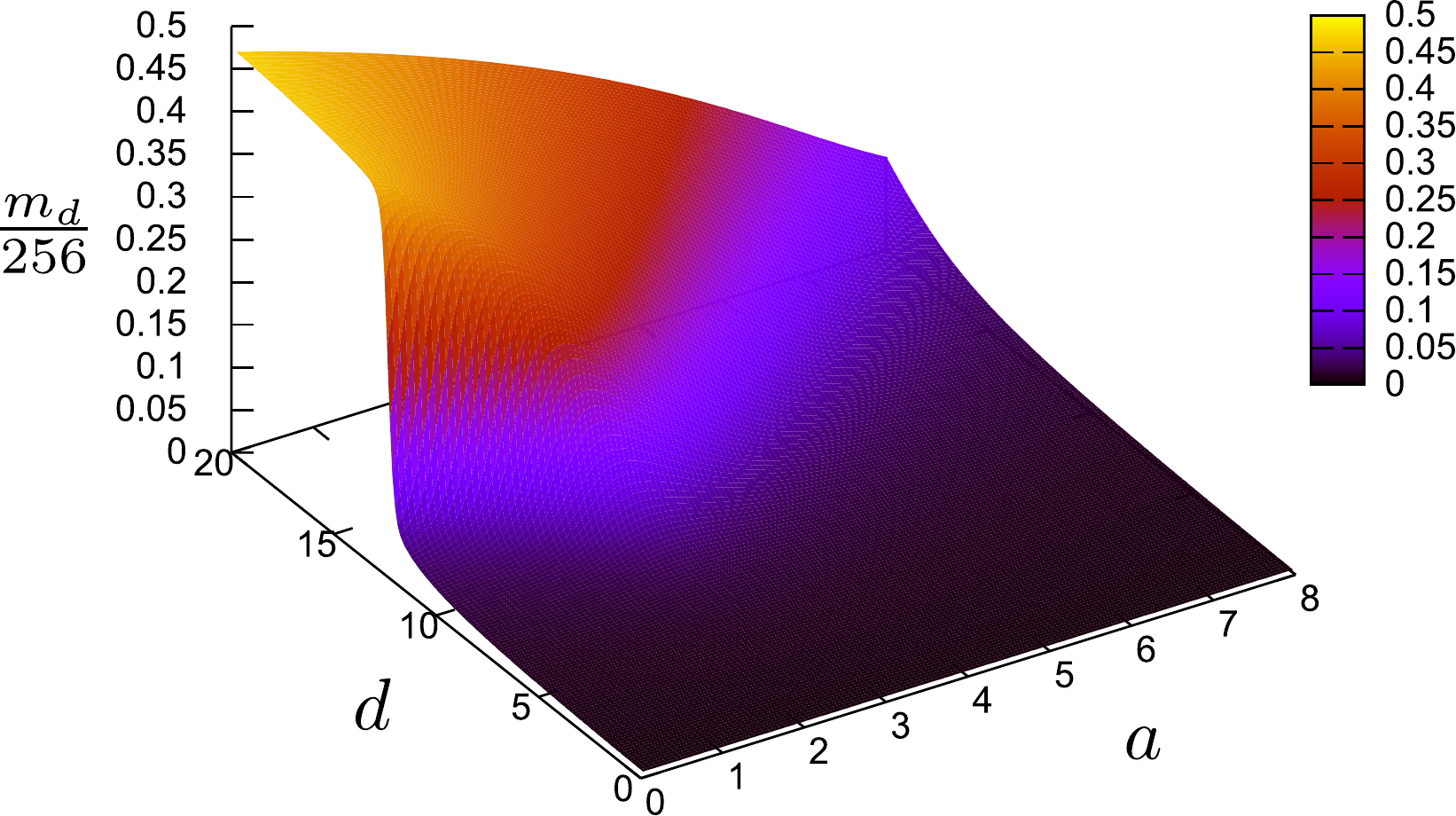}
\end{center}
\caption{A plot of the density of $d$ visits calculated at length $n=256$. This
highlights the region where it is tends to a non-zero constant and corresponds
well to the region where we have shown that $z_d$ is the dominant singularity.}
\label{md ord parameter}
\end{figure}

If we define 
\begin{align}
{\cal{A}}(a,d) &= \lim_{n\rightarrow \infty} \frac{\langle m_a \rangle}{n}
& \text{ and } &
& {\cal{D}}(a,d) & = \lim_{n\rightarrow \infty} \frac{\langle m_d \rangle}{n},
\end{align}
then in the desorbed phase we have 
\begin{align}
{\cal{A}}&={\cal{D}}=0,
\end{align}
while in the $a$-rich phase we have 
\begin{align}
{\cal{A}}&>0 
&\text{ and } &
&{\cal{D}}=0,
\end{align}
and in the $d$-rich phase has both 
\begin{align}
{\cal{A}}&>0 
&\text{ and } &
&{\cal{D}}>0.
\end{align}
In Figures~\ref{ma ord parameter} and ~\ref{md ord parameter} we plot
$\frac{\langle m_a \rangle}{256}$ and $\frac{\langle m_d \rangle}{256}$
respectively.

The phase boundary between the desorbed and $a$-rich phases
occurs at 
\begin{align}
  a &= 2 &\mbox{for $d<d_c(2)$}.
\end{align}
Note that this phase boundary is, unsurprisingly, independent of $d$. We can
compute $d_c(2)$ exactly using the results of Section~\ref{sec sol a012}:
\begin{align}
  G(2,2;1/4) &= \sum_{n \geq 0} C_n C_{n+1} 16^{-n} = 8 - \frac{64}{3\pi} ;\\
  d_c(2) &= \frac{2 G(2,2)}{G(2,2)-1} 
  = \frac{16(8-3\pi)}{64-21\pi} \approx 11.55159579.
\end{align}
In a similar way we can compute $d_c(0)$ and $d_c(1)$:
\begin{align}
  d_c(0) &= \frac{30\pi}{165\pi-512} \approx 14.81234030;\\
  d_c(1) &= \frac{8(512-165\pi)}{4096-1305\pi} \approx 13.47187382.
\end{align}

\begin{figure}[h]
\begin{center}
 \includegraphics[height=5cm]{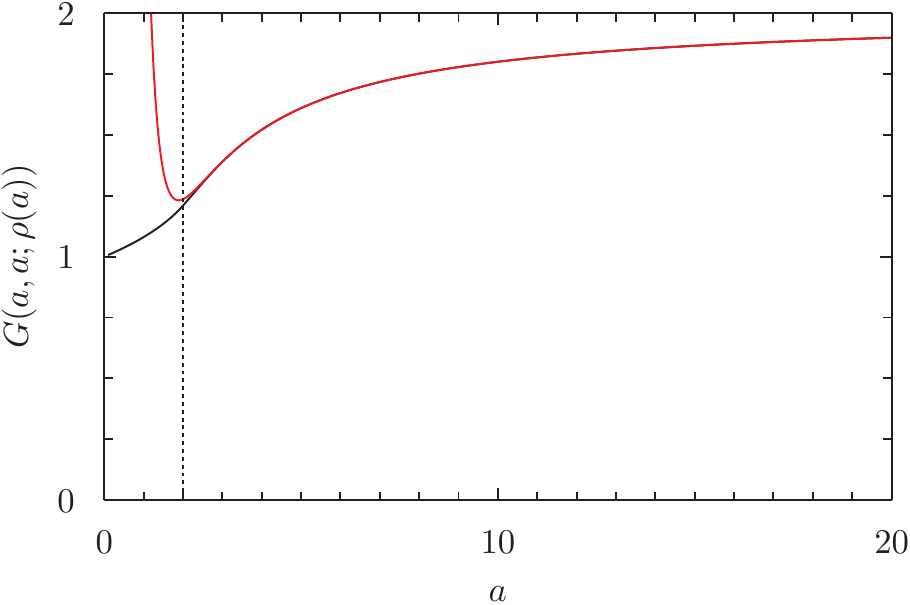}
 \includegraphics[height=5cm]{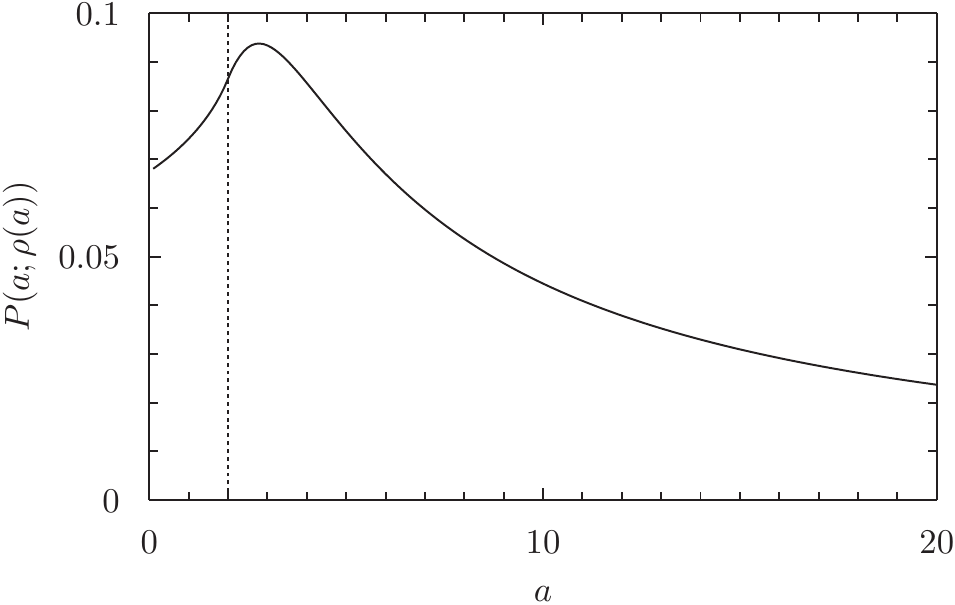}
\end{center}
\caption{A plot of both $G(a,a;\rho(a))$ and $P(a;\rho(a))$. The dotted line
indicates $a=2$. In the plot of $G(a,a;\rho(a))$ we have also marked the
asymptotic form computed below.}
\label{fig gaa pa}
\end{figure}

The transitions to the $d$-rich phase from the desorbed and $a$-rich phases are
given by
\begin{align}
  d_c(a) &= P(a;\rho(a))^{-1} = \frac{a G(a,a;\rho(a) )}{G(a,a;\rho(a))-1},
\label{eqn_arich-to-drich}
\end{align}
where $\rho(a)$ is given by equation~\Ref{eqn rhoa}. We plot $P(a;\rho(a))$ and
$G(a,a;\rho(a))$ in Figure~\ref{fig gaa pa}.

In the limit as $a \to \infty$, $d_c(a) \to 2a + o(a)$. As $a \to \infty$, the
generating function $G(a,a)$ is dominated by those configurations which have a
maximal number of visits. In this case, the lower walk simply zig-zags along
the wall and the upper walk is effectively unconstrained by the lower. Hence
\begin{align}
  \lim_{a \to \infty} G(a,a;z) &= \sum_{n \geq 0} C_n a^n z^{2n} =
\frac{1-\sqrt{1-4az^2}}{2az^2} .
\end{align}
Substituting $z = \rho(a)$ then gives $G(a,a;\rho(a)) = 2 + o(1)$ from which
the asymptotics of $d_c(a)$ follows:
\begin{equation}
d_c(a) \sim 2 \, a \qquad \mbox{ as } \qquad a \rightarrow \infty.
\end{equation}
In the next section we compute the above asymptotic form in more detail.

\begin{figure}[h]
\begin{center}
 \includegraphics[height=8cm]{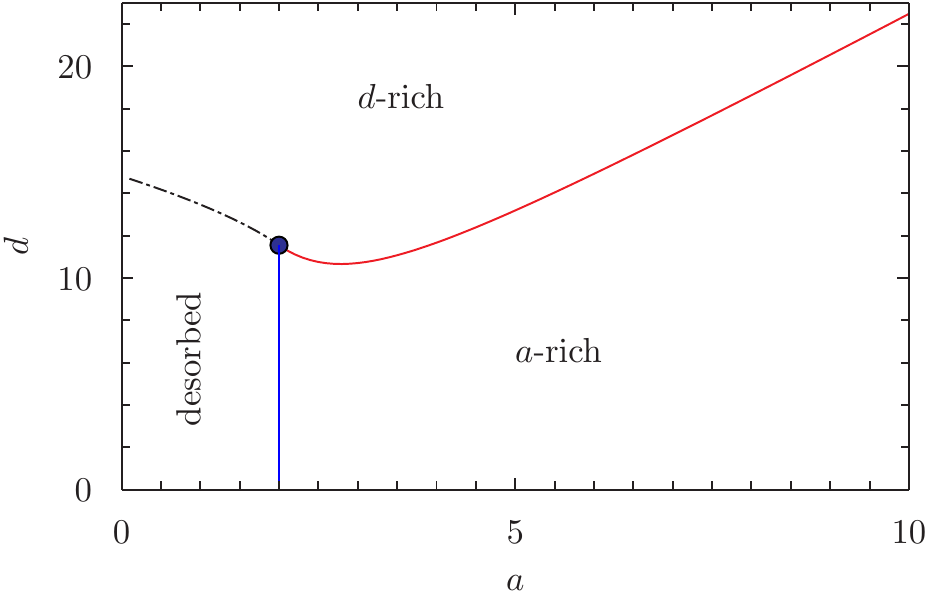}
\end{center}
\caption{The phase diagram of our model. The three phases are as indicated and
the first-order transition is marked with a dashed line, while the two
second-order transitions are marked with solid lines. The three boundaries meet
at the point $(a,d) = (2,11.55\dots)$. }
\label{fig boundaries}
\end{figure}

Combining all of this information gives the phase diagram for our model, which
we present in Figure~\ref{fig boundaries}. It is interesting to note that the
three transition lines meet with the two critical lines forming an angle.
Classically this would indicate mean-field like behaviour of a bicritical
point. If true, this mean-field behaviour would be interesting to understand. 

\subsection{Asymptotics of the $d$-rich-$a$-rich phase boundary}
We now consider how the $d$-rich $a$-rich phase boundary, $d_c(a)$, behaves for
large $a$ in more detail. From equation~\Ref{eqn Gaa GV}, we write $[z^{2n}
a^k]G(a,a;z)$ in closed
form
\begin{align}
  [z^{2n} a^k]G(a,a;z) &= 
  \frac{k(k+1)(k+2)}{(n+1)^2(n+2)(2n-k)} \binom{2n}{n} \binom{2n-k}{n}.
\end{align}
We seek the asymptotic form of $G(a,a;\rho(a))$ as $a \to \infty$, and so we
need to evaluate the asymptotics of 
\begin{align}
  G(a,a;\rho(a)) &= \sum_{n \geq 0} 
  \frac{(a-1)^n}{(n+1)^2(n+2)4^n a^{2n}} \binom{2n}{n}
  \sum_{k=0}^n \frac{k (k+1)(k+2)}{(2n-k)} \binom{2n-k}{n} a^k.
\end{align}
Expanding this slightly further gives
\begin{align}
  G(a,a;\rho(a)) &= 
  \sum_{n \geq 0} \frac{1}{(n+1)^2(n+2) 4^n a^{2n}} \binom{2n}{n}
  \sum_{j,k=0}^n \frac{(-1)^j k (k+1)(k+2) }{(2n-k)}
  \binom{n}{j} \binom{2n-k}{n} a^{j+k}. \\
\intertext{Substitute $j = \ell - k$ to get}
  &= 
  \sum_{n \geq 0} \frac{1}{(n+1)^2(n+2) 4^n} \binom{2n}{n}
  \sum_{\ell = 0}^{2n} a^{\ell-2n}
  \sum_{k=0}^n \frac{(-1)^{\ell-k} k(k+1)(k+2) }{(2n-k)}
  \binom{n}{\ell-k} \binom{2n-k}{n}.
\end{align}
So now the coefficient of $a^0$ is
\begin{align}
 [a^0] &= \sum_{n=0}^\infty \frac{1}{n+1}\binom{2n}{n} 4^{-n} = 2,\\
 \intertext{while}
 [a^{-1}] &= - \sum_{n=1}^\infty \frac{3n}{(n+1)(n+2)}\binom{2n}{n} 4^{-n} = -2,
\\
 [a^{-2}] &= 0 \\
 [a^{-3}] &= \sum_{n=2}^\infty \frac{2(n-1)}{(n+1)(n+2)}\binom{2n}{n} 4^{-n} =
1,\\
 \intertext{and}
 [a^{-4}] &= \sum_{n=3}^\infty \frac{3(n-2)}{(n+1)(n+2)}\binom{2n}{n} 4^{-n} =
\frac{5}{4}.
\end{align}
These simple forms continue as far as we have observed. This then gives
\begin{align}
 G(a,a;\rho(a)) &\sim 2 - \frac{2}{a} + \frac{0}{a^2} + \frac{1}{a^3} +
\frac{5}{4a^4} + \frac{15}{16 a^5} + \frac{7}{32a^6} + O(a^{-7}).
\end{align}
We can then plot this against our numerical estimates of $G(a,a;\rho(a))$. One
should note that the series $G(a,a,\rho(a))$ converges very slowly for $a>2$.
Since we know the summands decay as $n^{-3/2}$, we can assume that the partial
sums, $s_n$, grow as $A + Bn^{-1/2}$. We can then accelerate the convergence of
the series by estimating $A$ with the sequence
\begin{align}
 A_n &= s_n ( n + \sqrt{n}\sqrt{n-1} )
      - s_{n-1} (n-1 + \sqrt{n}\sqrt{n-1} ).
\end{align}
This combination was chosen by solving the pair of simultaneous equations $s_n =
A+Bn^{-1/2}, s_{n-1} = A + B(n-1)^{-1/2}$. We found that sequence $A_n$
converged far faster than the partial sums $s_n$.

\section{Discussion}
\label{sec:disc}

\subsection{Nature of solution}
\label{subsec:sol-nature}

In the section~\ref{sec:analpd} we demonstrated that when $a=1$, the model undergoes a phase
transition at $d=d_c(1) = \frac{8(512-165\pi)}{4096-1305\pi}$. Since this is not
an algebraic number, it follows that the generating function of the model does
not satisfy a linear differential equation in $z$ with integer polynomial
coefficients in $a,d$ and $z$. That is, it cannot be D-finite.

Consider, to the contrary, that the generating function $G(1,d;z) = [r^0
s^0]f(r,s;1,d;z)$ is a D-finite power series in $z$ with integer polynomial
coefficients in $d$. By definition, it satisfies a non-trivial linear
differential equation of the form
\begin{align}
  p_k(d;z) \ppdiff{k}{G}{z}+ \dots
  p_1(d;z) \pdiff{G}{z} + p_0 G(1,d;z) &= 0,
\end{align}
where the $p_j(d;z)$ are integer polynomials in $d$ and $z$.
By standard results in the theory of linear differential equations, the
singularities of $G(1,d;z)$ are zeros of the leading polynomial $p_k(d;z)$. 

For small $d$ we know  that  the dominant singularity of $G(1,d)$ is
$z_b=1/4$. At the critical value of $d$, there is a change in dominant
singularity from $z_b$ to $z_d$. Exactly at the critical value, $z_b =
z_d = 1/4$. Thus the discriminant of $p_k(d;z)$ with respect to $d$ must be zero
at this point. Since the discriminant is a polynomial in $z,d$ with integer
coefficients and $z=1/4$, it follows that this critical value of $d$ must be an
algebraic number. Above we showed that $d_c(1)=
\frac{8(512-165\pi)}{4096-1305\pi}$ which is not algebraic and thus $G(1,d;z)$
is not D-finite. A standard result \cite{lipshitz1989a-a} on D-finite series
states that specialisation of D-finite series are themselves D-finite and thus 
$G(a,d;z)$ cannot be D-finite and nor is $f(r,s;a,d;z)$.

The Lindstr\"om-Gessel-Viennot lemma combines the partition functions of
single-walk models --- equivalent to sums and Hadamard products of the
underlying single-walk generating function which is algebraic (this is true
quite generally --- see \cite{bousquet2008a-a}). Any finite
combination of Hadamard products and sums of algebraic or D-finite generating
functions remains D-finite \cite{lipshitz1989a-a} and thus the
Lindstr\"om-Gessel-Viennot lemma (alone) cannot be applied to decompose the
model considered here into single-walk problems.

That being said, the Lindstr\"om-Gessel-Viennot lemma can be combined with a
factorisation argument to yield a solution as we demonstrated in section~\ref{sec:altsol}.

\subsection{Fixed energy ratio models: $r$-models}
\label{subsec:rmod}
Finally, let us now consider the family of physical models parameterised by $-\infty < r < \infty$ where
\begin{align}
\varepsilon_d &= r \varepsilon_a
& \text{and so} &&
d &= a^r
\end{align}
that allows us to summarise our results. Let us call these $r$-models.

For any $r$-model the high temperature phase is the desorbed state. The model
effectively already analysed by Brak \emph{et\ al.}\ \cite{brak1998c-:a,owczarek2001c-:a} has $d=a$ and so $r=1$. In this case there
is a single low temperature phase being the $a$-rich phase. Given there
are no additional phase boundaries for $a<d$ one can deduce that for all
$r\leq1$ the model goes from the desorbed state at high temperatures through a
single second-order phase transition to the $a$-rich phase at low temperatures.

The special point in our phase diagram where $(a,d)=(2,11.55\ldots)$ where the
three phases meet occurs in the $r$-model with
\begin{equation}
r = r_t\equiv \frac{\log(11.55\ldots)}{\log{2}} = 3.53\ldots .
\end{equation}
For all $r\geq r_t$ there is a single low temperature phase which is the
$d$-rich phase: the transition on lowering the temperature is now first-order.

\begin{figure}[h]
\begin{center}
 \includegraphics[height=8cm]{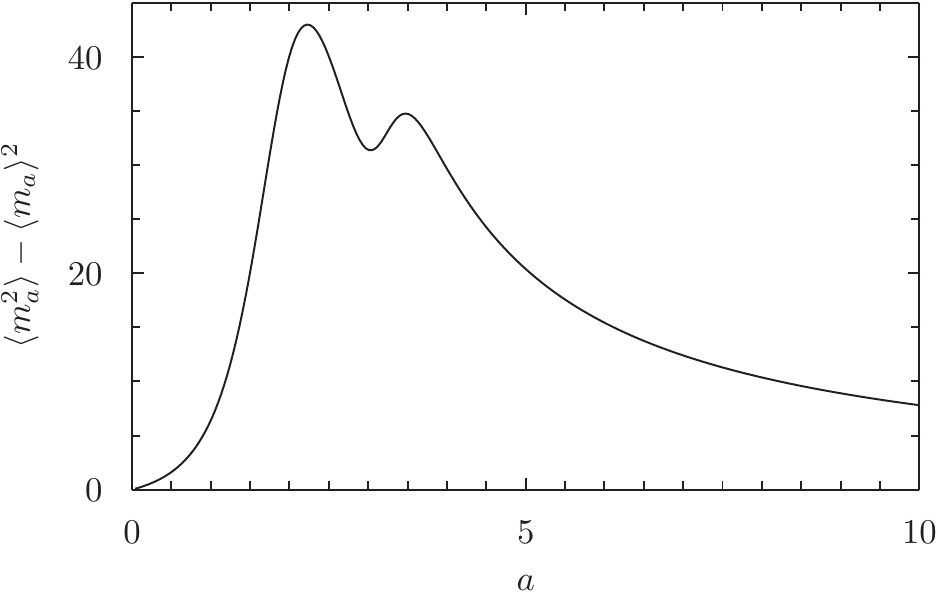}
\end{center}
\caption{A plot of the fluctuations in the number of $a$-visits, $m_a$, for
length $n=128$ as function of $a$ clearly showing two peaks.}
\label{fluctuations r2}
\end{figure}

Since we have shown above that $d_c(a) \sim 2a$ as $a \rightarrow \infty$ one
can now argue that for all $1< r < r_t$ the $r$-model has \emph{two} phase
transitions on lowering the temperature. At very low temperatures the model is
in a $d$-rich phase while at high temperatures the model is in the desorbed
state. At intermediate temperatures the system is in an $a$-rich phase. Both
transitions, from desorbed to $a$-rich, and $a$-rich to $d$-rich, are
second-order transitions with jump discontinuities in the specific heat. In
Figure~\ref{fluctuations r2} we plot the fluctuations in $a$-visits as a
function of temperature at length 128 for the $r=2$ $r$-model: two peaks occur
in these fluctuations.

If one argues that a physically realisable model is one where both walks pick up
the same energy when they touch the surface together then  the model is the one
with $r=2$, that is $d=a^2$. It is interesting to see that this model contains
two phase transitions: one at $a=2$ and the other at $a\approx 3.301$
found by solving equation~(\ref{eqn_arich-to-drich}) for $d_c=a_c^2$. In any
case, we have a family of adsorption models that have one or two low temperature
states and which the order of the transition changes as the parameter is varied.
We have analysed this model using an exact solution and fully delineated its
behaviour.  It will be of interest to analyse the behaviour of this model in a
slit.

\section*{Acknowledgments}
Financial support from the Australian Research Council via its support
for the Centre of Excellence for Mathematics and Statistics of Complex
Systems and the Discovery Projects scheme is gratefully acknowledged by one of
the authors, ALO. We thank the referees for their comments and questions.
Finally, AR and ALO thank Yao-ban Chan for his comments on the manuscript and
Sophie for her latt\'e art.

%\bibliography{refs-aleks-general,refs-aleks-personal}
%\bibliographystyle{aip}

\end{document}